\documentclass[preprint,aps]{revtex4}
\usepackage{epsfig}
\begin{document}
\title{Quarks, Electrons, and Atoms in Closely Related Universes }
\author{Craig J. Hogan}
\address{Astronomy and Physics Departments, 
University of Washington,
Seattle, Washington 98195-1580}
\begin{abstract}
In a model where a multiverse wavefunction explores a multitude of vacua with different symmetries and parameters,     properties of universes closely related to ours can be understood by examining the consequences of  small departures of physical parameters from their observed values. The masses of the light fermions that make up the stable matter of which we are made--- the up and down quarks, and the electron--- have values in a narrow window that both allows a variety of   nuclei other than protons to exist, and at the same time allows atoms with stable shells of electrons that are not devoured by their nuclei. These fundamental parameters of the Standard Model are good candidates for quantities whose values  are determined through  selection effects within a multiverse, since a   living world of  molecules needs stable nuclei other than just protons and neutrons. If the fermion  masses are fixed  by brane condensation or compactification of extra dimensions, there may be observable fossils of the branching event, such as a gravitational wave background.

\end{abstract}
\maketitle
\section{Introduction}

We know that nature is governed by mathematics and symmetries.
Not very long ago, it was an article of faith among most  physicists that everything about physics would eventually be explained in terms of fundamental symmetries--- that nothing in the makeup of physical laws is accidental, that nature ultimately has no choices, and that all the properties of particles and fields  are fixed by pure math. 

In the thirty  years since modern anthropic reasoning was introduced into cosmology\cite{carter,carr}, the competing idea that   anthropic  selection might have an indispensable role in fundamental physical theory has gradually become, if not universally accepted, at least mainstream. There are now concrete physical models for realizing anthropic selection in nature. Cosmology has provided not only a concrete mechanism (inflation) for manufacturing multiple universes, but also a new  phenomenon (Dark Energy) whose value is most often explained by invoking anthropic explanations. String theory has uncovered a  framework by which many different symmetries and parameters  for   fields can be realized in the low-energy, 3+1-dimensional universe, depending on the topology and size of the  manifold of the other seven, truly fundamental dimensions, and on the configurations of p-branes within it, especially the local environment of the 3-brane on which our own Standard Model fields live. The numbers of locally metastable configurations of   manifold and branes, and therefore the number of options for low energy physics, are estimated to be so large that for all practical purposes, there is a continuum of choices for fundamental parameters that we observe\cite{Bousso:2000xa,Kachru:2003aw,Susskind:2003kw}.

Of course, the details of how this works in the real world are still  sketchy. Cosmology unfolds in a series of phase transitions and symmetry breakings. For example, it is now part of standard inflation that the    quantum wavefunction of the universe branches early into various options for the zero-point fluctuations of the inflaton field, different branches of which correspond to different distributions of galaxies. String theory opens up a scenario in which    the multiverse wavefunction
 may also branch  very early into a variety of whole universes, each of which has  different physics. If things happen this way, it is natural for us to find ourselves in a branch with physics remarkably well suited to make the stuff of which we are made. 
 
 It then makes sense to ask new questions about the world: how  would things change if this or that aspect of physics were changed? If a small change in a certain parameter changes the world a great deal in a way that matters to our presence here, that is a clue that that particular parameter  is fixed by selection rather than by symmetry.  The following arguments along these lines are elaborated more fully in ref. \cite{Hogan:1999wh}.
 
 Now we may be faced where a situation where some seemingly fundamental  features of physics might not ever be derived from first principles. Even the particular gauge group  in our Grand Unified Theory (that is, the the one in our branch of the wavefunction) might be only one  group selected out of many options provided by the Theory of Everything. We may have to adjust our scientific style to this larger  physical reality, which forces cosmology and fundamental physics into a new relationship. For example, although  we can't look inside the other universes of the multiverse ensemble and can't predict the branching outcome from first principles,   cosmological experiments now under development might reveal relict gravitational waves from the same symmetry breaking that fixed the parameters.

\section{Changing Standard Model  Parameters}

Evaluating changes in the world in response to changes in the fundamental physics is actually a  difficult program to carry out. For the most fundamental theory we have, the Standard Model, the connection of many of  its parameters with generally observable phenomena can only be roughly estimated. First-principles calculations of the behavior of systems such as nuclei and molecules are possible only for the simplest examples.

The traditional minimal  Standard Model has 19 ``adjustable'' parameters\cite{cahn,gaillard}: Yukawa coefficients
fixing the  masses of the six quark
and three lepton flavors ($u,d,c,s,t,b,e,\mu,\tau$), the Higgs mass
and vacuum expectation value $v$ (which multiplies the
Yukawa coefficients to determine the fermion masses), three angles and one
phase of the CKM (Cabibbo-Kobayashi-Maskawa) matrix (which mixes quark weak- and
strong-interaction eigenstates), a phase for the quantum chromodynamic (QCD) vacuum,
and three coupling constants $g_1,g_2,g_3$ of the  
gauge group, $U(1)\times SU(2)\times SU(3)$.
We now know experimentally that the neutrinos are not massless, so there
 are at least seven more parameters to characterize their behavior
(three masses and another four CKM matrix elements). Thus 26 parameters, plus Newton's constant $G$ and the cosmological constant $\Lambda$ of General Relativity,  are enough to describe the behavior  of all the observed 
particles in all   experiments, except those related to new Dark Matter particles.
 If in addition the Standard Model is extended by supersymmetry, the number of parameters exceeds 100.  

Imagine that you are sitting at a control panel of the universe. It has   a few dozen knobs--- one for each of the parameters. Suppose you start twiddling the knobs. For all  but a few of the knobs, you find nothing changes very much; the mass of the top quark for example (that is, its Yukawa coupling coefficient in the Standard Model equations)  has little direct effect on everyday stuff. Which knobs matter for the stuff we care about most--- atoms and molecules?

Some knobs are clearly important, but their exact value does not seem too critical. The fine structure constant $\alpha$ for example controls the sizes of all the atoms and molecules, scaling like the Bohr radius $(\alpha m_e)^{-1}$.  If you twiddle  this knob, natural phenomena dominated by this physics---  which includes all of familiar chemistry and   biology--- grow or shrink in size. On the other hand they all grow or shrink by almost the same fractional amount so  the structural effect of changes is  hard to notice; the miraculous fit of base pairs into the DNA double helix would still work pretty well. There are however   subtle changes in structural relationships and molecular reaction rates. Your complicated biochemistry   probably would not survive a sudden big change in $\alpha$, but if you turn  the knob slowly enough, living things probably adapt to the changing physics.  Simulations of cellular reaction networks show that their behavior is remarkably robust with respect to changes in reaction rates, and mostly depend on network topology\cite{munro}.

It turns out that a few  of the knobs have a particularly large qualitative  effect with  a very small amount of twiddling. Three knobs stand out for their particularly conspicuous effects:  the Yukawa coefficients controlling the  masses of the electron,  the  up-quark, and the down-quark. They are the  light  fermions that dominate the composition and behavior of atoms and molecules. Changing them by even a small fractional amount has a devastating effect on whether  molecules can exist at all. The most dramatic sensitivity of the world on their values seems to be in the physics of atomic nuclei.

\section{Effects of changing $u,d,e$ masses on atoms and nuclei}

The light fermion masses are all very small (less than one percent) compared with the mass of even a single proton. (Protons and neutrons, which comprise the bulk of the mass of ordinary matter, ironically have a mass dominated not by  the ``real mass'' of their matter particles, the constituent  quarks, but almost entirely by the kinetic energies of the quarks and the massless gluons mediating the color forces.) However, the light fermion masses are critical because they determine the energy thresholds for reactions that control the stability of nucleons.

In the three dimensional parameter space formed by these masses,  the most reliable phenomemological statements can be made about changes within  the two dimensional surface defined by holding the sum of $u$ and $d$ masses constant. (That is because many complicated features of nuclear physics remain constant if the pion mass, which is proportional to $(m_u+m_d)^{1/2}$, is constant). In this plane, 
some properties of worlds with different values of the masses  are summarized in figures 1 and 2, the latter taken from ref. \cite{Hogan:1999wh}. The figures also show  a constraint from a particular SO(10) grand unified scenario, to illustrate that likely unification schemes probably do not leave all these parameters independent--- at least one relationship between them is likely fixed by unification symmetry. 

In the lower part of  figure 1, towards larger  up-quark mass, there are ``Neutron Worlds''.  As one dials knobs in this direction, a threshold is soon crossed  where it is energetically favorable for the electron in a hydrogen atom to join with its proton to make a neutron. If you turn it farther, even a free proton (without any nearby electron) spontaneously decays into a neutron. 

In the upper part of the figure, there are ``Proton Worlds''. Moving up from our world, a threshold is soon crossed where a deuteron in a plasma is no longer energetically favored over a pair of protons.  If you go farther, even an isolated deuteron spontaneously decays into a pair of protons. 

In the neutron world, there are  nuclei,  but not atoms with electrons around them, so   chemistry does not  happen. In the proton world, there are hydrogen atoms, but that is the only kind of atom, because the other nuclei do not  form or are not stable. Fortunately for us there is a world in between, where a few dozen stable nuclei are both possible and are actually produced in stars, and are endowed with  electron orbitals leading to  chemistry with arbitrarily large and complex molecules. This world would disappear  with only a few percent fractional change in the quark mass difference in either direction. It does not exist in some closely-related branches of the multiverse wavefunction.

\begin{figure} 
\epsfysize=5in 
\epsfbox{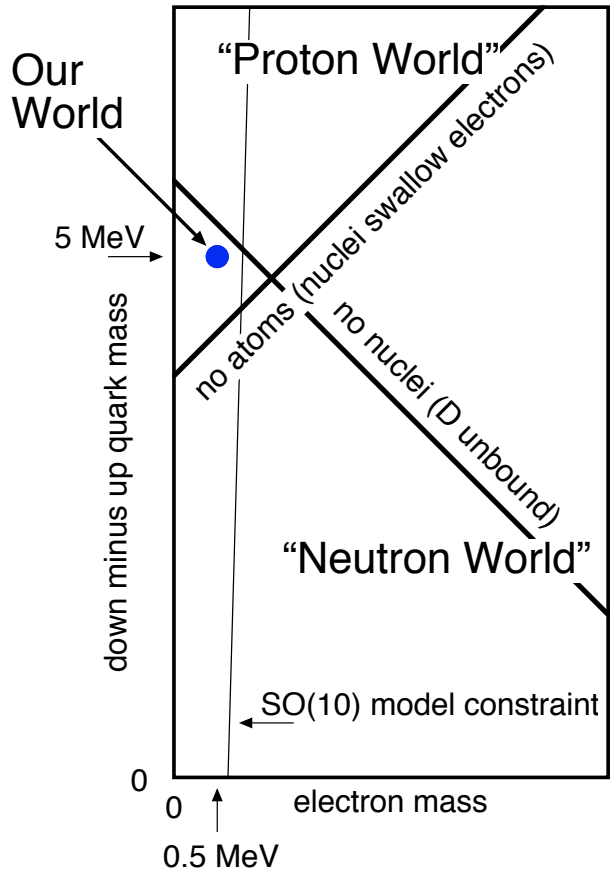} 
\caption{ An overview of simple nuclear physics of the neutron, proton and deuteron,  in other universes closely related to ours. Thresholds for various reactions are shown depending on the mass difference   between the down and up quark mass, and  the electron mass, in the plane where the sum of the up and down masses does not change.  Our world is the [pale blue] dot. The SO(10) constraint shown imposes the restriction that the ratio of electron to down quark mass is fixed by a symmetry to have the same value it does in the real world; the region to the right of this is excluded for positive down-quark mass.}
\end{figure}

\begin{figure} 
\epsfysize=3.5in 
\epsfbox{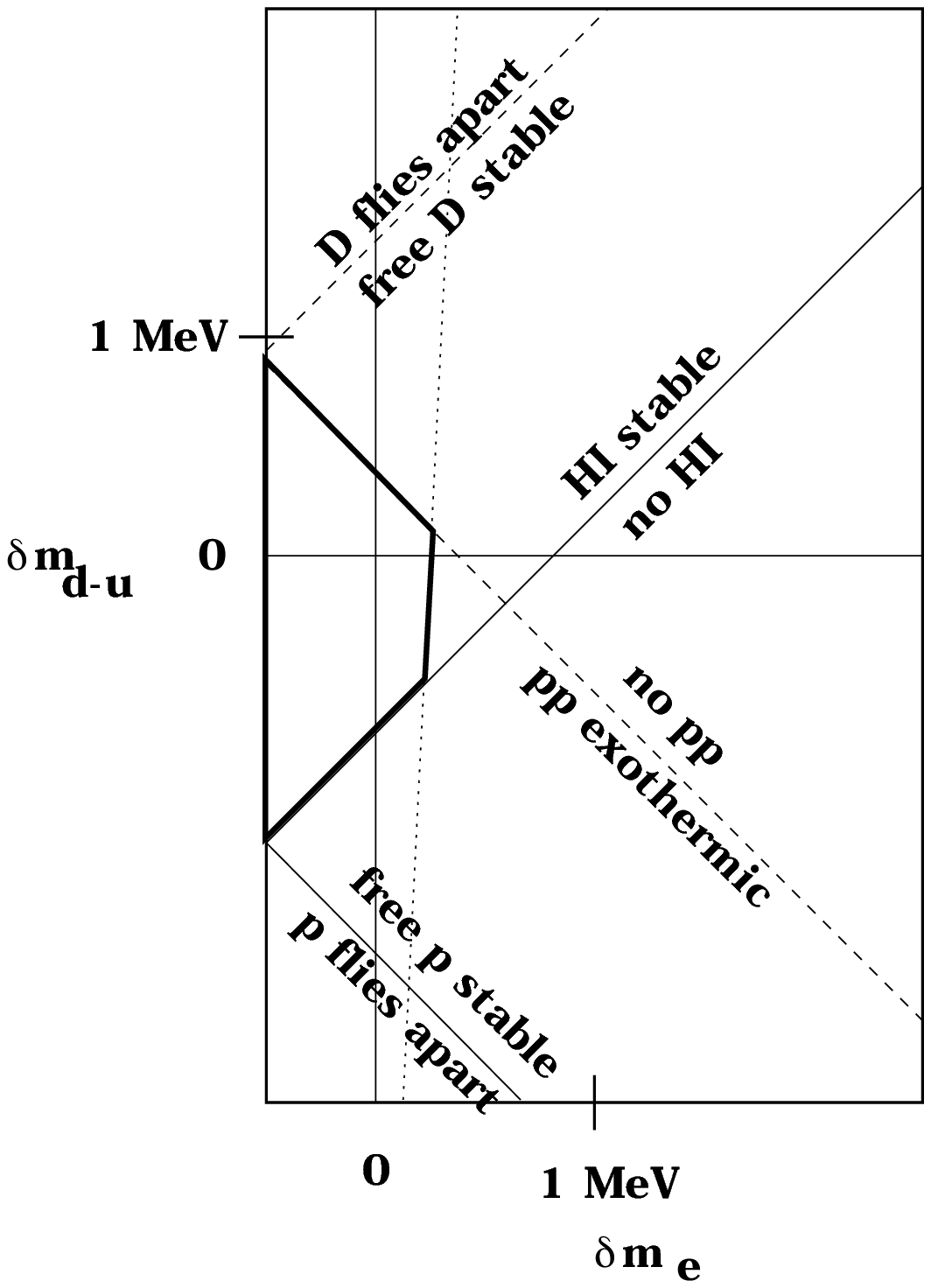} 
\caption{\label{fig: fermionsold} A more detailed view, from ref. \protect\cite{Hogan:1999wh}, of the changes in   thresholds of nuclear reactions, as a function of the change in the $u,d$ mass difference and the change in the electron mass. Our world is at the origin in these quantities. }
\end{figure}

One can   estimate roughly  the effects of leaving this plane. In that case, nuclear physics is changed in new ways,  since the mass of the pion  changes.    It appears that if  the masses are increased by more than about 40\%, the range of nuclear forces is reduced to the point where the deuteron is unstable; and if they are reduced by a similar amount, they are strengthened to the point where the diproton is stable. On the other hand, the latter change also reduces the  range of nuclear forces   so there are fewer stable elements overall. The sum of the quark masses in our world appears roughly optimized for the largest number of stable nuclei.  Again, the  situation would change qualitatively (e.g., far fewer stable elements) with changes in summed quark masses at the ten percent level.

Why is it even possible to find parameters balanced between the neutron world and the proton world? For example, if the SO(10) model is right one, it seems that  we are lucky that its trajectory passes through the region that allows for molecules.
The answer could  be that even the gauge symmetries and particle content   also have an anthropic explanation. A great variety of compact 7-manifolds and 3-brane configurations solve the fundamental M theory. Each one of them has dimensional scales corresponding to parameter values such as particle masses, as well as topological and geometrical relationships corresponding to symmetries. Many of these configurations undergo inflation and produce macroscopic universes. In this situation it is not surprising that we find ourselves in one  where atoms and nuclei can exist.

\section{quantum mechanics of anthropic selection}

Discussions of anthropic selection have sometimes differentiated between the kind that selects whole universes (with different values of the electron mass, etc.), and the kind that selects a congenial environment (why we do not live on an asteroid or a quasar, etc.) While these seem very different, from a quantum-mechanical perspective they do not differ in kind.  Both of them are selections of a congenial branch of the wavefunction of the universe. 

In the original formulation of quantum mechanics, it was said that an observation collapsed a wavefunction to one of the eigenstates of the observed quantity. The modern view is that the cosmic wavefunction never collapses, but only appears to collapse from the point of view of observers who are part of the wavefunction. When Schr\"odinger's cat lives or dies, the branch of the wavefunction with the dead cat also contains observers who are dealing with a dead cat, and the branch with the live cat  also contains observers who are petting a live one.

\begin{figure} 
\epsfysize=3in 
\epsfbox{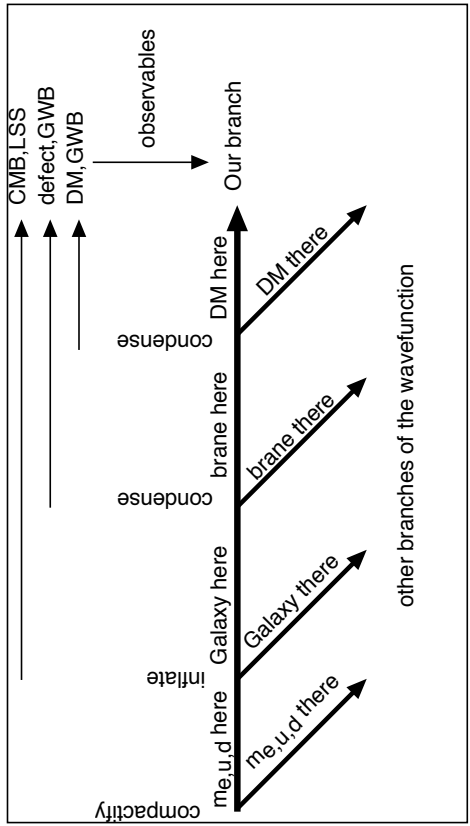} 
\vskip 1.5in
\caption{ A schematic sketch of the branching history of the wavefunction to which we belong. At various points in cosmic history, symmetry breaking  (e.g.,  compactification, inflation, condensation) made random choices,  which were frozen into features such as standard model parameters, the galaxy distribution,  or the dark matter density. In some cases, these events left other observables which can be observed directly in other ways, such as microwave background anisotropy, large scale structure, gravitational wave backgrounds, or cosmic defects. Thus although the other branches of the wavefunction cannot be observed directly, the physics of the branching events in some situations may be independently observable.}
\end{figure}

Although this is sometimes called the ``Many Worlds'' interpretation of quantum mechanics, it is really about having just one world, one wavefunction, obeying the Schr\"odinger equation: the wavefunction evolves linearly from one time to the next based on its previous state. Anthropic selection in this sense is built into physics at the most basic level of quantum mechanics. Selection of a wavefunction branch is what drives us into circumstances   in which we thrive. Viewed from a disinterested perspective outside the universe, it   looks like living beings swim like salmon up their favorite branches of the wavefunction, chasing their favorite places. 

The selection of a planet or a galaxy is  a matter of chance. In quantum mechanics this means   a branch of the wavefunction has been selected. The binding energy of our galaxy was determined by an inflaton fluctuation during inflation; that was when the branching occurred that selected the large scale gravitational potential that set the parameters for our local cosmic environment.  We can achieve statistical understanding about this kind of selection because we can observe other parts of the ensemble, by observing galaxy clustering, the microwave background, and so on. In this way, we understand the physics of the symmetry breaking.  We even know something about the formation of the different galaxy distributions in other universes we will never see. These are regarded as just different by chance.

If the quark and  electron masses are also matter of chance, the branching of the wavefunction   occurred along with  the symmetry breaking  that fixed their masses. There may be ways to observe aspects of the statistical ensemble for this event also, by 
studying the gravitational wave background rather than the microwave background.

We do not know when all the choices of parameters and symmetries are made. Some of these branchings may leave  traces of other choices observable in our past light cone. It could be that some parameters are spatially varying even today, in response to spatial variations in scalar or dark matter fields. For example, one model of dark energy predicts   large variations in  the masses of neutrinos, depending on the local density  of the neutrino component of  dark matter\cite{Fardon:2003eh}. 
(Indeed the basic idea that effective neutrino masses depend on the local physical environment is now part of the standard theory of solar neutrino oscillations.)  Thus the properties of stars can be spatially modulated depending on the dark matter density--- a quantity  determined, in  many theories, by a branching event that occurred recently enough to have an observable ensemble.  Such ideas provide a new motivation for observational programs to quantify the extent to which the constants of nature are really constant in spacetime. (A thriving example of this can be found in studies of varying $\alpha$.)

In some models, events connected with fixing the local quark and electron masses  may have happened   late enough to leave fossil traces. This could happen during   the final compactification of some of the extra dimensions, or the condensation of our own Standard Model 3D brane within a larger dimensional space.

 If  dimensional compactification happens in a sufficiently catastrophic symmetry breaking,  it  can lead to a background of gravitational waves. Because they are so penetrating, gravitational waves  can carry information directly from almost the edge of our past light cone,  well  beyond recombination, even beyond weak decoupling--- indeed, back to the edge of 3D space as we know it.   If the extra dimensions are smaller than the Hubble length at dimensional compactification or brane condensation, their collapse  can appear as a first-order phase transition in our 3D space, leading to  relativistic flows of mass-energy.  If the extra dimensions are larger than or comparable to the Hubble scale, the    3D brane we inhabit  may  initially condense with warps and wiggles that lead to a gravitational wave background. Either way the
mesoscopic,  classical motion of branes settling down to their final equilibrium configuration could lead to a strong gravitational-radiation background in a frequency range detectable by detectors now under development\cite{Hogan:2000aa,Hogan:2000is,Hogan:2001jn,Ichiki:2004sx}. Thus, instruments designed to observe the early boundary of spacetime may also explore the early boundary of physics as we know it, and directly test ideas concerning the separation of various branches of the multiverse having different fundamental parameters. 

This blending of empirical cosmology and fundamental physics is reminiscent of our 
Darwinian   understanding of the tree of life. The double helix, the four-base codon alphabet and the triplet genetic code for amino acids, any particular gene for a protein in a particular organism---   are all frozen accidents of evolutionary history. 
It is futile to try to understand or explain these aspects of life, or indeed any relationships in biology,  without referring to the way the history of life unfolded.  In the same way that  (in Dobzhansky's phrase), ``nothing in biology  makes sense except in the light of evolution'',     physics in these models only makes  sense in the light of cosmology.

\begin{acknowledgements}
This work was supported by NSF grant AST-0098557 at the University of
Washington.

\end{acknowledgements}

{}

\end{document}